\documentclass{article}[12pt]
\usepackage[T1,T2A]{fontenc}
\usepackage[cp1251]{inputenc}
\usepackage{amsmath,amssymb}
\usepackage{graphicx}%
\usepackage{multirow}

\newtheorem{stat}{Statement}

\newcommand{\Fig}[3]{%
\begin{center}
\parbox{8cm}{%
\refstepcounter{figure}\includegraphics[width=8cm,height=#2cm]{#1} \noindent Fig. \thefigure:\quad
#3}\end{center}}

\newcommand{\ThreeFigReg}[9]{%
\begin{flushleft}
\begin{tabular}{lcr}
\hspace{-1cm}
\parbox{6.5cm}{\includegraphics[width=6.5cm,height=#2cm]{#1}}  & \hspace{-1cm} \parbox{6.5cm}{\includegraphics[width=6.5cm,height=#4cm]{#3}}  & \hspace{-1.5cm}\parbox{6.5cm}{\includegraphics[width=6.5cm,height=#6cm]{#5}}\\
\hspace{-10pt}
\parbox{5cm}{\vspace{7pt}\refstepcounter{figure}Figure \thefigure.\quad #7\vfill} &\hskip -1cm \parbox{5cm}{\vspace{7pt}\refstepcounter{figure}Figure \thefigure.\quad #8\vfill} & \hspace{-1cm}\parbox{5cm}{\vspace{7pt}
\refstepcounter{figure}Figure \thefigure.\quad #9\vfill}\\
\end{tabular}
\end{flushleft}
\vspace{7pt}
}

\newcommand{\TwoFigsReg}[7]{%
\begin{flushleft}
\begin{tabular}{cc}
\parbox{#3cm}{\centerline{\includegraphics[width=#3cm,height=#4cm]{#1}}}  & \parbox{#3cm}{\centerline{\includegraphics[width=#3cm,height=#5cm]{#2}}}  \\
\parbox{#3cm}{\vspace{7pt}\refstepcounter{figure}Figure \thefigure.\quad #6\vfill} & \parbox{#3cm}{\vspace{7pt}\refstepcounter{figure}Figure \thefigure.\quad #7\vfill}\\
\end{tabular}
\end{flushleft}
\vspace{7pt}
}

\newcounter{strochka}

\newcounter{spisok}
\begin{document}

\begin{center}
{\bf \Large Study of a complete model of the cosmological evolution of a classical scalar field with a Higgs potential.
IV. Large-scale model transformations.
\\[12pt]
Yu. G. Ignat’ev and A. R. Samigullina}\\[12pt]
Institute of Physics, Kazan Federal University, Kremlyovskaya str., 16A, Kazan, 420008, Russia
\end{center}

\begin{abstract}
A study and numerical modeling of the cosmological evolution of a classical scalar field with the Higgs potential was carried out. Based on the formulated similarity properties of cosmological models, their main characteristics are studied for models with different interaction scales: the Planck scale, the Grand Unified scale and the Standard Model scale. Based on numerical integration, the similarity properties of these models are proven with high accuracy. \\

{\bf keywords}: cosmological models, Higgs fields, Einstein-Higgs hypersurfaces, similarity properties of dynamical systems, numerical modeling.\\
{\bf PACS}: 04.20.Cv, 98.80.Cq, 96.50.S 52.27.Ny
\end{abstract}

\section{Introduction}
As is known, Higgs scalar fields \cite{Hig} were introduced into particle physics in connection with the development of the so-called Standard Model (see, for example, \cite{Vai}), which was experimentally confirmed at the Hadron Collider \cite{JHEP} . In turn, scalar fields were introduced into cosmology in the 80s of the 20th century in connection with the need to explain the inflationary expansion of the Universe (see, for example, \cite{Gor}). In the work \cite{Bel}, an analysis of the model of the Universe based on a classical scalar field with a quadratic potential was carried out using methods of the qualitative theory of dynamical systems (see, for example, \cite{Bog}). In the work \cite{Zhu} a similar analysis was carried out for a two-component cosmological model, consisting of a scalar field and a neutral ideal fluid.
Subsequently, many studies were carried out on the qualitative and numerical analysis of cosmological models with various forms of potential energy. Unfortunately, the bulk of these works pursued the goal of obtaining exact solutions of Einstein’s equations, according to which the form of potential energy was selected, often not related to specific field-theoretic models of elementary particles. In the works \cite{Ign1}-\cite{Ign3} a qualitative and numerical analysis of the cosmological model based on the classical Higgs scalar field was carried out. In these works, features of the behavior of cosmological models were discovered and studied depending on the values of their fundamental parameters. Unfortunately, technical difficulties did not allow us to carry out full-fledged numerical simulations for real values of these parameters. In \cite{Ign4} a number of properties of mathematical models based on scalar fields and charged fermions were proven with respect to the similarity transformation, allowing the results of numerical simulations to be extended from one model to another similar to it. In this article, based on the invariant properties of similarity, we will conduct a more complete study of models with real values of fundamental parameters.

\section{Mathematical model}
\subsection{General equations of the model}

As a field model, we consider a self-consistent system of equations of Einstein and the classical scalar field $\Phi$ with the Higgs potential, which corresponds to the Lagrange function
\begin{equation} \label{Eq1}
L_s=\frac{1}{16\pi}\bigg(g^{ik} \Phi _{,i} \Phi _{,k} -2V(\Phi )\bigg),
\end{equation}
  where $V(\Phi)$ is the potential energy of the scalar field:
\begin{equation} \label{Eq2}
V(\Phi)=\displaystyle\frac{\alpha}{4}\left(\Phi^2-\frac{m^2}{\alpha}\right)^2,
\end{equation}
  $\alpha$ is the self-interaction constant, m is the mass of scalar bosons. The Lagrange function \eqref{Eq1} corresponds to the scalar field equation:
\begin{equation} \label{Eq3}
\Delta\Phi+V'_\Phi=0,
\end{equation}
where $\Delta=g^{ik}\nabla_i\nabla_k$ , and the energy - momentum tensor of the scalar field
\begin{equation} \label{Eq4}
    T^i_k=\frac{1}{16\pi}\left(2\Phi^{,i}\Phi _{,k}-\delta^i_k\Phi _{,j}\Phi^{,j }+2V(\Phi)\delta^{i}_k\right)
   \end{equation}
The corresponding Einstein equations of the system under study have the form\footnote{Everywhere $G=\hbar=c=1$ , metric signature (-,-,-,+), the Ricci tensor is obtained by convolution of the first and third indices.}:
\begin{equation} \label{Eq5}
G_k^i\equiv R^i_k-\frac{1}{2}\delta^i_kR=8\pi T^i_k+\delta^i_k\Lambda_0.
   \end{equation}
where $\Lambda_0$ is the initial value of the cosmological constant, associated with its observed value $\Lambda$, obtained by removing the constant term in the potential energy, by the relation:
\begin{equation} \label{Eq6}
\Lambda=\Lambda_0-\frac{m^4}{4\alpha}.
\end{equation}
Equations \eqref{Eq3} and \eqref{Eq5} are the basic equations of the model.

\subsection{Equations of the comological model}

In the case of a spatially flat Friedmann Universe
\begin{equation} \label{Eq7}
  ds^2_0=dt^2-a^2(t)(dx^2+dy^2+dz^2),
\end{equation}
where $a(t)$ is the scale factor. Let us introduce the Hubble parameter
\begin{equation} \label{Eq8}
H=\displaystyle\frac{\dot{a}}{a}
\end{equation}
In this case $\Phi=\Phi(t)$, and the energy-momentum tensor of the scalar field takes on an isotropic structure
\begin{equation} \label{Eq9}
T^i_k=(\varepsilon+p)\delta_4^i\delta_k^4-p\delta^i_k,
\end{equation}
Where
\begin{equation} \label{Eq10}
\varepsilon=\frac{1}{8\pi}\left(\frac{\dot{\Phi}}{2}+V(\Phi)\right); \text{ } p=\frac{1}{8\pi}\left(\frac{\dot{\Phi}}{2}-V(\Phi)\right)
\end{equation}
is the energy density of the scalar field, so
\begin{equation*}
\varepsilon+p=\frac{1}{8\pi}\dot{\Phi}^2
\end{equation*}
Thus, the system of equations \eqref{Eq3} - \eqref{Eq5} with respect to the scale factor $a(t)$ and scalar potential $\Phi(t)$ is reduced to an autonomous dynamic system:
\begin{equation} \label{Eq11}
\dot{\Phi}=Z(\equiv F_2)
\end{equation}
\begin{equation} \label{Eq12}
\dot{H}=-\frac{Z^2}{2}(\equiv F_1\leq 0)
\end{equation}
\begin{equation} \label{Eq13}
\dot{Z}=-3HZ-m^2\Phi+\alpha\Phi^3 (\equiv F_3)
\end{equation}
Note that the Einstein equation for the $^4_4$ component is the first integral of the system \eqref{Eq11} - \eqref{Eq13}:

\begin{equation} \label{Eq14}
\Sigma:3H^2-\frac{Z^2}{2}-\frac{m^2\Phi^2}{2}+\frac{\alpha\Phi^4}{4}-\Lambda
\end{equation}
Equation \eqref{Eq14} defines some 4th order algebraic surface in the phase space $\mathbb{R}^3$ of the dynamical system $\Sigma\subset \mathbb{R}^3$, on which all phase trajectories of the dynamical system lie , i.e., the history of specific cosmological models. In what follows we will call the surface $\Sigma$ \emph{Einstein-Higgs hypersurface}. Equation \eqref{Eq14} determines the initial value of the Hubble parameter $H(0)\equiv H_0$ for given initial values $\Phi(0)\equiv \Phi_0$ and $Z(0)\equiv Z_0$. Two symmetric solutions for the initial value of the Hubble parameter $H_0^{\pm}\equiv \pm H_0$ correspond to starting from the expansion state (+) or from the compression state (-).
The equation \eqref{Eq14} can be written in the form
\begin{equation*}
3H^2-\mathcal{E}=0,
\end{equation*}
where the non-negative quantity $\mathcal{E}$ is \emph{effective energy of the system}:
\begin{equation} \label{Eq15}
\mathcal{E}(\Phi,Z)\equiv\frac{\dot{\Phi}^2}{2}+\frac{m^2\Phi^2}{2}-\frac{\alpha\ Phi^4}{4}+\Lambda\geq 0
\end{equation}
The autonomous system \eqref{Eq11} - \eqref{Eq13} is invariant with respect to time translations $t\rightarrow t_0+t$ . Thus, when choosing the sign of the initial value of the Hubble parameter, only two initial values remain free: $\Phi_0$ and $Z_0$, which we will specify as an ordered list:
\begin{equation} \label{Eq16}
I=[\Phi_0,Z_0,\nu], \text{} (\Phi_0=\Phi(0), Z_0=Z(0)),
\end{equation}
where $\nu=\pm 1$ , and the value +1 corresponds to the start of the system from the state $H_0>0$, and the value -1 to the start from the state $H_0<0$. In this case, regions of the phase space $V\subset \mathbb{R}^3$ in which the energy condition \eqref{Eq15} is violated, i.e.,
\begin{equation} \label{Eq17}
V\subset \mathbb{R}^3:\mathcal{E}(\Phi,Z)<0,
\end{equation}
are inaccessible to a dynamical system, as a result of which the phase space of a dynamical system may turn out to be multiply connected.

Note that according to \eqref{Eq12} the Hubble parameter in this model can only decrease with time, and $H=\text{Const} \Leftrightarrow \dot{\Phi}=0$ . Introducing also \emph{invariant cosmological acceleration}
\begin{equation} \label{Eq18}
\Omega\equiv\frac{a\ddot{a}}{\dot{a}^2}\equiv 1+\frac{\dot{H}}{H^2},
\end{equation}
Let's present the equation \eqref{Eq12} in an equivalent form:
\begin{equation} \label{Eq19}
\Omega=1-\frac{1}{2}\frac{Z^2}{H^2}(\leq 1).
\end{equation}
Let us also introduce a useful relation for \emph{quadratic invariant of space curvature} Friedman:
\begin{equation} \label{Eq20}
\sigma\equiv\sqrt{R_{ijkl}R^{ijkl}}=H^2\sqrt{6(1+\Omega^2)}\geq 0.
\end{equation}

So, the dynamic system under study has 3 degrees of freedom, and its state is uniquely determined by the coordinates of the point $M(\Phi,Z,H)\equiv M(x_1,x_2,x_3)$ in the 3-dimensional phase space $\mathbb{R }^3$ . Note that inaccessible regions \eqref{Eq17}, if they exist, are cylindrical with axes parallel to $OH$.

\subsection{Scaling model transformations}

So, the cosmological model under study is a three-dimensional dynamical system in the arithmetic phase space $\mathbb{R}^3=\{\Phi,Z,H\}$ and is completely described by an autonomous system of first-order ordinary differential equations \eqref{Eq11} - \ eqref{Eq13} with the integral condition \eqref{Eq14}. In turn, the system of equations \eqref{Eq11} - \eqref{Eq14} is determined by an ordered set of fundamental constants
\begin{equation} \label{Eq21}
\textbf{P}=[\alpha,m,\Lambda]
\end{equation}
Further, the geometric locus of the phase space points $M(\Phi,Z,H)\equiv[\Phi(t),Z(t),H(t)]$ determines the phase trajectory corresponding to the specific history of the cosmological model.
Since the system of equations \eqref{Eq11} - \eqref{Eq13} is essentially nonlinear, only its qualitative analysis is possible. For a detailed study of a specific history, numerical integration of these equations is necessary. As preliminary studies have shown, the numerical integration of this system in the region of sufficiently small values of the fundamental constants \eqref{Eq21}, which is of physical interest, encounters significant technical difficulties that do not allow us to trace the evolution of the system over sufficiently large cosmological times. In order to get around these difficulties, consider large-scale transformations of the system \eqref{Eq11} - \eqref{Eq13} \cite{Bog}:
\begin{eqnarray}\label{Eq22}
\begin{array}{l}
\alpha=k^2\tilde{k},\Lambda=k^2\tilde{\Lambda},m=k\tilde{m},t=k^{-1}\tilde{t}\\
\Phi(t)=\tilde{\Phi}(kt),Z(t)=k\tilde{Z}(kt)),H(t)=k\tilde{H}(kt)\Leftrightarrow V(\Phi)=k^2\tilde{V}(\tilde{\Phi})
\end{array}
\end{eqnarray}
The following statement is true (see \cite{Bog}):

\begin{stat}
Scale transformations \eqref{Eq22} leave the system \eqref{Eq11} - \eqref{Eq13} invariant.\label{St1}
\end{stat}

Note that the invariant acceleration $\Omega$ \eqref{Eq18}, and together with it the quadratic invariant of space curvature $\sigma$ \eqref{Eq20} and the coefficient $\kappa (p=\kappa\mathcal{E})$ barotropes are also invariant under the transformation \eqref{Eq22}
\begin{equation} \label{Eq23}
\Omega(t)=\tilde{\Omega}(kt),\sigma(t)=\tilde{\sigma}(kt),\kappa(t)=\tilde{\kappa}(kt)
\end{equation}
and the expression for the effective energy $\mathcal{E}$ \eqref{Eq15} is transformed according to the law
$$\mathcal{E}\rightarrow k^2\mathcal{E}$$
Note that the equation \eqref{Eq14} of the Higgs hypersurface, which determines the topology of phase trajectories, is invariant with respect to scale transformations \eqref{Eq22}. Therefore, the topology of this surface does not change during scale transformations \eqref{Eq22}.

\subsection{Invariants of the matrix of a dynamical system}
The main matrix $\mathbf{A}$ of the dynamical system \eqref{Eq11} - \eqref{Eq13} (see, for example, \cite{Bog}) is
\begin{eqnarray}\label{Eq24}
\mathbf{A}=
\left(\begin{array}{ccc}
0&1&0\\
-m^2+3\alpha\Phi^2&-3H&-3Z\\
m^2\Phi-\alpha\Phi^3&0&-6H
\end{array}\right).
\end{eqnarray}
The singular points of a dynamical system are determined by the equality of the right-hand sides of the normal system of equations to zero, and the eigenvalues are determined by the eigenvalues of this matrix at the singular points. During scale transformations \eqref{Eq22} the right sides of the dynamic equations \eqref{Eq11} - \eqref{Eq13} are transformed according to the rule:
$$\tilde{F}_1=\frac{1}{k}F_1,\tilde{F}_2=\frac{1}{k^2}F_2,\tilde{F}_3=\frac{1} {k^2}F_3$$
Therefore, the characteristic matrix \eqref{Eq24} of the dynamic system is transformed according to the rule:

\begin{eqnarray}\label{Eq25}
\mathbf{\tilde{A}}=
\left(\begin{array}{ccc}
0&1&0\\
&&\\
\displaystyle\frac{-m^2+3\alpha\Phi^2}{k^2}&\displaystyle\frac{-3H}{k}&\displaystyle\frac{-3Z}{k}\\
&&\\
\displaystyle\frac{m^2\Phi-\alpha\Phi^3}{k^2}&0&\displaystyle\frac{-6H}{k}
\end{array}\right).
\end{eqnarray}
The eigenvalues $\lambda$ and eigenvectors $\mathbf{X}$ of this matrix are defined by the equation
\begin{eqnarray*}
\begin{array}{c}
\mathbf{AX}=\lambda\mathbf{X},
\end{array}
\mathbf{X}=\left(\begin{array}{c}
\Phi\\
Z\\
H
\end{array}\right).
\end{eqnarray*}
The images $\tilde{\lambda}, \tilde{X}$ are determined by the corresponding equations.
\begin{eqnarray}\label{Eq26}
\begin{array}{c}
\mathbf{\tilde{A}\tilde{X}}=\lambda\mathbf{\tilde{X}},
\end{array}
\mathbf{\tilde{X}}=\left(\begin{array}{c}
\Phi\\
Z/k\\
H/k
\end{array}\right).
\end{eqnarray}
Carrying out the calculations, we come to the following conclusion:

\begin{stat}
During scaling transformations \eqref{Eq22} the eigenvalues of the characteristic matrix of the dynamic system \eqref{Eq11} - \eqref{Eq13} are transformed according to the law.\label{St2}
\end{stat}
\begin{equation} \label{Eq27}
\tilde{\lambda}=\frac{\lambda}{k}
\end{equation}

Statements \ref{St1} and \ref{St2} establish an important property of the dynamic system under study - e\"e \emph{property of mechanical similarity}. Thanks to the property of similarity, it is possible to transfer the results of the study of a specific dynamic system to an infinite number of similar systems connected by transformations \eqref{Eq22} This factor becomes fundamentally important in numerical modeling, when the numerical integration of dynamic equations in the case of too small or too large values of the parameters \textbf{P} \eqref{Eq21} becomes difficult to implement when studying the behavior of the system at large times (at the study of these models has to extend numerical integration to times of the order of $t\sim 10^{17}\div 10^{18}$ We will use this property of similarity of a dynamical system in this article.The fact is that this system of equations is quite simple is numerically integrated for values of the fundamental constants \emph{P} of the order of unity $\alpha \sim 1, m\sim 1, \Lambda \sim 1$ and not for very large values of time $t$. The physically interesting range of parameter values corresponds to values of the fundamental parameters that are several orders of magnitude less than those indicated.

In what follows we will consider three main field-theoretical models:
\begin{eqnarray}\label{Eq28}
\begin{array}{ll}
M_0:\alpha \sim 1, m\sim 1, \Lambda \sim 0.1 & \text{ -- base model;}
\end{array}
\end{eqnarray}
\begin{eqnarray}\label{Eq29}
\begin{array}{ll}
M_{SU(5)}:\alpha \sim 10^{-10}, m\sim 10^{-5}, & \text{ -- SU(5)model}\vspace{5pt}\\
  \Lambda \sim 10^{-11}\Rightarrow k=10^5 &\text{Grand Unified\footnote{Modern versions of the Grand Unified theory include masses of Higgs bosons even on the order of $10^{17}$ GeV};}
\end{array}
\end{eqnarray}
\begin{eqnarray}\label{Eq30}
\begin{array}{ll}
M_{SM}:\alpha \sim 10^{-30}, m\sim 10^{-15}, & \vspace{5pt}\\
  \Lambda \sim 10^{-31}\Rightarrow k=10^{15} &\text{ -- standard model.}
\end{array}
\end{eqnarray}
\subsection{Singular points of a dynamical system}
Singular points of a dynamical system, as well as the geometry of the Higgs hypersurface, depending on the fundamental parameters, were studied in detail in the work \cite{Ign1}. Below we briefly list the main results needed in this article.
Thus, the coordinates of the singular points of the dynamic system \eqref{Eq11} - \eqref{Eq13} are determined by the equations:
\begin{eqnarray}\label{Eq31}
\begin{array}{l}
Z=0;-3HZ-m^2\Phi=\alpha\Phi^3=0\\
-3H^2+\displaystyle\frac{m^2\Phi^2}{2}-\displaystyle\frac{\alpha\Phi^4}{4}+\Lambda=0
\end{array}
\end{eqnarray}
Hence, in the general case, we have 6 singular points of the system - two symmetrical with zero potential and its derivative \cite{Ign1}:
\begin{eqnarray}\label{Eq32}
\begin{array}{ll}
M_{\pm}\left(0,0,\pm \sqrt{\displaystyle\frac{\Lambda}{3}}\right),&(\text{if } \Lambda\geq 0,\forall \alpha )
\end{array}
\end{eqnarray}
  and 4 symmetric $M_{ab}$, located at the vertices of the rectangle in the plane $P_2\{\Phi,H\}$: $Z=0$
\begin{eqnarray}\label{Eq33}
\begin{array}{lll}
M_{ab}\left(\pm \displaystyle\frac{1}{\sqrt{m\alpha}},0,\pm \sqrt{\displaystyle\frac{\Lambda_\alpha}{3}}\right) ,&\text{if }\alpha>0&\Lambda_\alpha \equiv \Lambda+\displaystyle\frac{1}{4\alpha}\geq 0
\end{array}
\end{eqnarray}
Thus, there are a total of 5 possible cases of \cite{Ign1}:
\begin{enumerate}
\item for $\{\alpha<0, \Lambda<0\}$ there are no singular points;
\item for $\{\alpha>0, \Lambda<-1/4\alpha\}$ there are no singular points;
\item for $\{\alpha<0, \Lambda>0\}$ there are only two singular points $M_\pm$;
\item for $\{\alpha>0, -1/4\alpha<\Lambda<0\}$ there are 4 singular points $M_1,M_2,M_3,M_4$;
\item for $\{\alpha>0, \Lambda>0\}$ all 6 singular points exist.
\end{enumerate}
Calculating the value of the effective energy \eqref{Eq21} at singular points, we find:
\begin{eqnarray}\label{Eq34}
\begin{array}{ll}
\mathcal{E}(M_\pm)=\Lambda; &\mathcal{E}(M_{ab})=\Lambda_\alpha.
\end{array}
\end{eqnarray}
?
Thus, from the conditions \eqref{Eq32} and \eqref{Eq33} it follows: \emph{all singular points of the dynamical system, if they exist, are accessible}. At singular points, the eigenvalues of the matrix of the dynamical system are equal to \cite{Ign1}:
\begin{eqnarray*}
\begin{array}{ll}
M_\pm:k_1=\mp2\sqrt{3\Lambda},&k_{2,3}=\mp\displaystyle\frac{1}{2}\sqrt{3\Lambda}\pm\displaystyle\frac{1 }{2}\sqrt{3\Lambda-4};\vspace{10pt}\\
M_{ab}:k_1=\mp2\sqrt{3\Lambda_\alpha};&k_{2,3}=\mp\displaystyle\frac{1}{2}\sqrt{3\Lambda_\alpha}\pm\ displaystyle\frac{1}{2}\sqrt{3\Lambda_\alpha-8}
\end{array}
\end{eqnarray*}
In pairs of symmetric points $(M_{11},M_{21})$ and $(M_{12},M_{22})$ the eigenvalues of the matrix of the dynamical system coincide. The signs in front of the first and second radicals in these formulas take on meanings independent of each other: the signs in front of the first radicals correspond to different pairs of points, the signs in front of the second radicals correspond to different eigenvalues. In this case, it is important to remember the necessary conditions for the existence of singular points \eqref{Eq32} and \eqref{Eq33}.

Thus, the singular points $M_\pm$ for $3\Lambda-4>0$ are unstable (saddle) points, and for $3\Lambda-4<0$ they are stable (attracting centers). At these points the Hubble constant is $H=\pm\sqrt{\Lambda/3}$ for $\Lambda>0$, and there is no scalar field. According to the formulas \eqref{Eq18} and \eqref{Eq20} the invariant cosmological acceleration and invariant curvature are equal:
  \begin{eqnarray}
\begin{array}{ll}\label{Eq35}
\Omega(M_\pm)=1;&\sigma(M_\pm)=\displaystyle\frac{2}{\sqrt{3}}\Lambda
\end{array}
\end{eqnarray}
Thus, the singular points $M_\pm$ correspond to inflationary solutions with a positive or negative Hubble constant: $a\sim e^{\pm\sqrt{\Lambda/3t}}$ . For $e=+1$, all eigenvalues of the matrix of a dynamical system at its singular points $M_{ab}$ are real and have different signs. Thus, for $e=+1$ all singular points are saddle points. For $e=-1$ and $\Lambda_\alpha<8/3$, the eigenvalues of the dynamic matrix at its singular points $M_{ab}$ become complex conjugate, and among them there are values corresponding to attraction.
?

\section{Numerical modeling}
Below are the results of numerical simulations for some of the most interesting cases, corresponding to different qualitative behavior of phase trajectories.

\subsection{$\mathbf{P_0}=[[-1,1],-0.1];\mathbf{I_0}=[1,0,1]$}
Figures \ref{Fig1}-\ref{Fig9} show the results of numerical simulations for the $\mathbf{P_0}$ model.
As can be seen from the graphs in Figures \ref{Fig1}-\ref{Fig3}, the phase tractors of the models,
like their Eistein-Higgs surfaces, they obey the laws of scaling transformation \eqref{Eq22}:
in Fig. \ref{Fig2} the scales of the Z and H axes are stretched $k=10^5$ times, and in Fig. \ref{Fig3} - $k=10^{15}$ times.
\ThreeFigReg{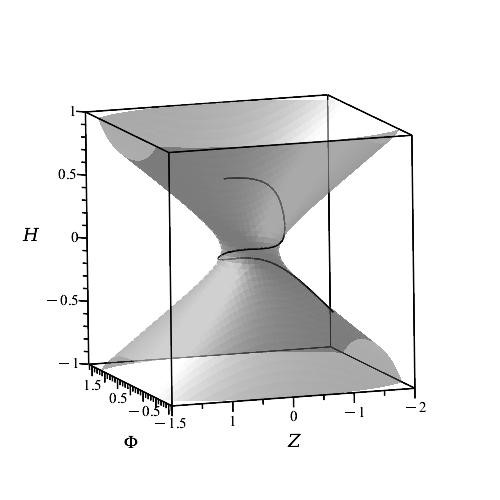}{6}{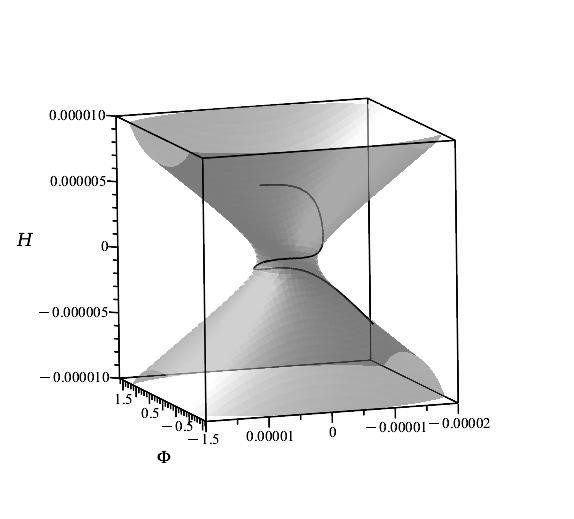}{6}{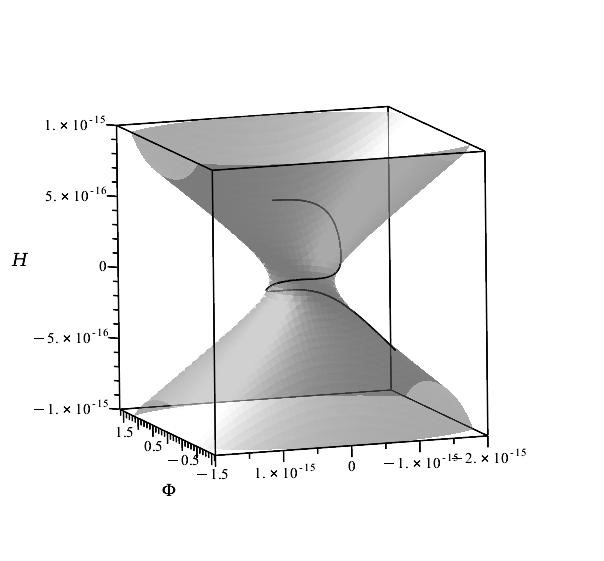}{6}{ \label{Fig1}Phase trajectory of the $\mathbf{M_0}$ model on the surface \eqref{Eq14}.}
{\label{Fig2}Phase trajectory of the model $\mathbf{M_{\mathrm{SU(5)}}}$ on the surface \eqref{Eq14}.}
{\label{Fig3}Phase trajectory of the model $\mathbf{M_{\mathrm{SM}}}$ on the surface \eqref{Eq14}.}
Figures \ref{Fig4} - \ref{Fig6} show graphs of the basic functions of the model $H(t)$ and $Z(t)$. On these graphs, the corresponding functions are presented on time scales corresponding to the transformations \eqref{Eq22}, i.e., for the SU(5) model $\tilde{t}=10^{-5}t$ , and for the SM model $ \ tilde{t}=10^{-15}t$ .
\TwoFigsReg{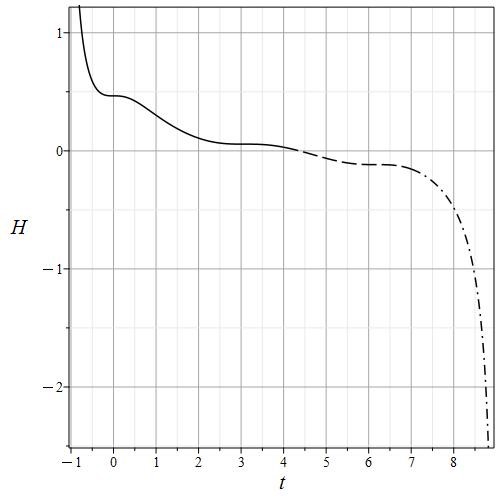}{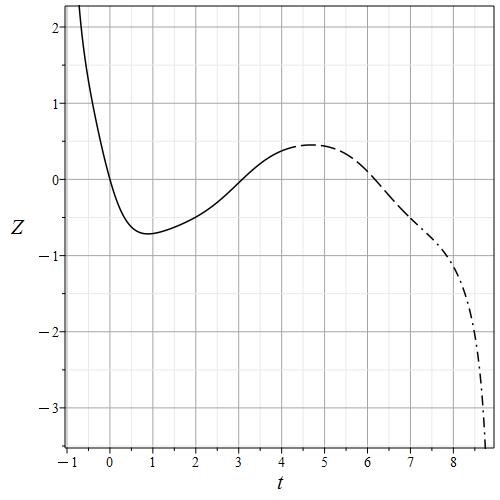}{8}{6}{6}{H(t): solid line -- $\mathrm{H_0}$; dashed -- $10^5\cdot H_{\mathrm{SU(5)}}$; dash-dotted - $10^{15}\cdot H_{\mathrm{SM}}$.\label{Fig4}}{Z(t): solid line -- $Z_0$; dashed - $10^{5}\cdot Z_{\mathrm{SU(5)}}$ ; dash-dotted -- $10^{15}\cdot Z_{\mathrm{SM}}$.\label{Fig5}}

As can be seen from the presented graphs, they completely obey the law of scaling transformation \eqref{Eq22}, their deviation from the prototype does not exceed $10^{-6}$, i.e., does not exceed the error of numerical integration.
Figures \ref{Fig6}-\ref{Fig7} show graphs of the invariant cosmological acceleration $\Omega(t)$ \eqref{Eq18} for these models, confirming its invariance (formula \eqref{Eq23}). However, we note that near the point $t_*\approx 4.36525$ the errors become significant, which is caused by the zero values of H(t) (see Fig. \ref{Fig4}) in the denominator of the formula \eqref{Eq18}.
Graphs in Fig. \ref{Fig6}-\ref{Fig9} precisely demonstrate the fact that the non-coincidence of the results near the singular point $t_*:H(t_*)=0$ is caused by insufficiently high calculation accuracy. These figures show similar graphs obtained by numerical integration with higher absolute $\Delta$ and relative $\delta$ integration accuracies.
\TwoFigsReg{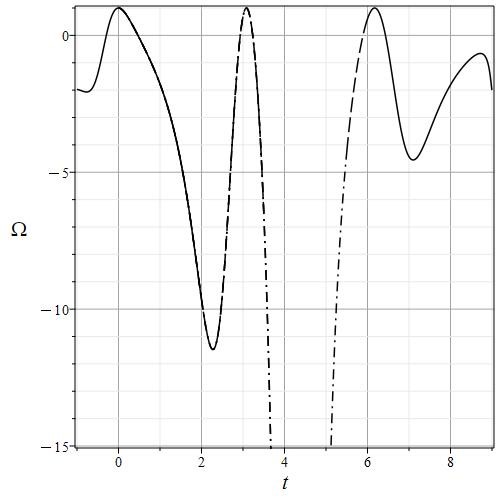}{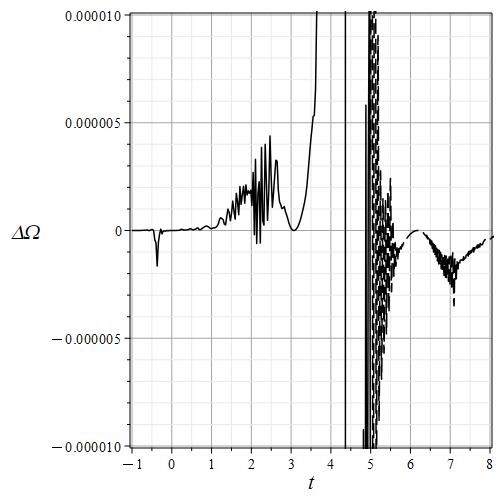}{8}{6}{6}{Invariant cosmological acceleration $\Omega(t)$:
$\mathrm{\mathbf{M_0}}$ -- solid line,
$\mathrm{\mathbf{M_{SU(5)}}}$ -- dash-dotted line,
$\mathrm{\mathbf{M_{SM}}}$ -- dotted point\label{Fig6}}
{Difference of cosmological accelerations $\Delta\Omega$:
$\Delta\Omega_{\mathrm{SU(5)}}\equiv\Omega_0-\Omega_{\mathrm{SU(5)}}$ -- solid line,
$\Delta\Omega_\mathrm{SM}\equiv\Omega_0-\Omega_{\mathrm{SM}}$ -- dash-dotted, with $\Delta=\delta=10^{-8}$\label{Fig7 }}
\TwoFigsReg{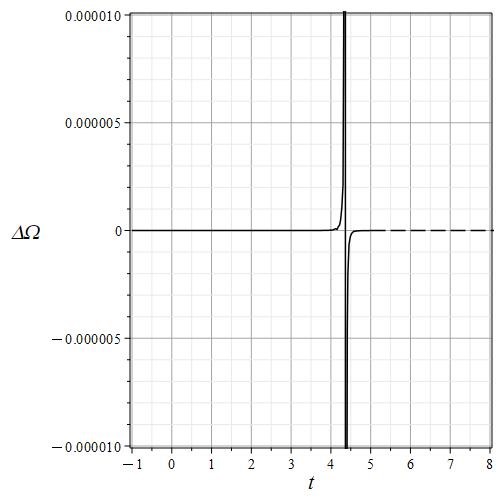}{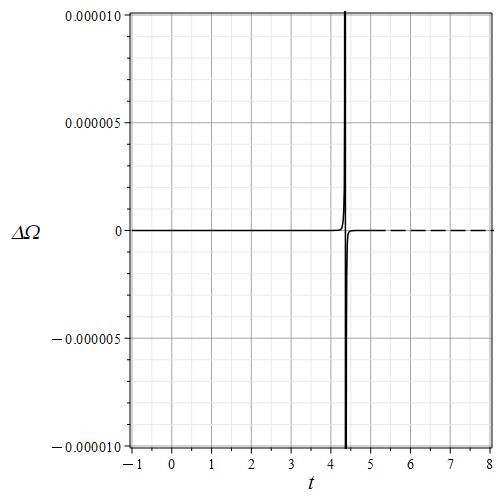}{8}{6}{6}{Difference of cosmological accelerations $\Delta\Omega$:
$\Delta\Omega_{\mathrm{SU(5)}}\equiv\Omega_0-\Omega_\mathrm{SU(5)}$- solid line,
$\Delta\Omega_{\mathrm{SM}}\equiv\Omega_0-\Omega_\mathrm{SM}$ - dash-dotted, with $\Delta=\delta=10^{-12}$\label{Fig8} .}
{Difference of cosmological accelerations $\Delta\Omega$:
$\Delta\Omega_{\mathrm{SU(5)}}\equiv\Omega_0-\Omega_\mathrm{SU(5)}$- solid line,
$\Delta\Omega_{\mathrm{SM}}\equiv\Omega_0-\Omega_\mathrm{SM}$ - dash-dotted, with $\Delta=\delta=10^{-13}$\label{Fig9} .}
\subsection{$\mathbf{P_1}=[[1,1],-0.3];\mathbf{I_1}=[0,0.8,1]$}
In this case, the Einstein-Higgs surface becomes doubly connected and there are no singular points on it. Figures \ref{Fig10} - \ref{Fig15} show the results of numerical simulations for the model
$\mathbf{P_1}$. These graphs exhibit similar properties with respect to scale transformations, as in the previous case,
although the type of trajectory becomes completely different. As can be seen from the graphs in figures \ref{Fig10}-\ref{Fig12},
phase tractors of models, as well as their Eistein-Higgs surfaces, obey the laws of scaling transformation \eqref{Eq18}:
in Fig. \ref{Fig11} the scales of the Z and H axes are stretched $k=10^5$ times, and in Fig. \ref{Fig12} - $k=10^{15}$ times.
\ThreeFigReg{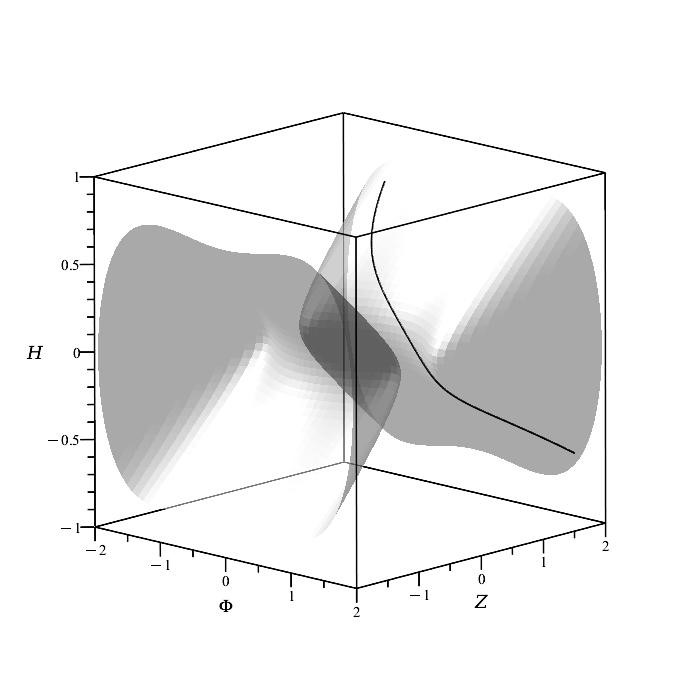}{6}{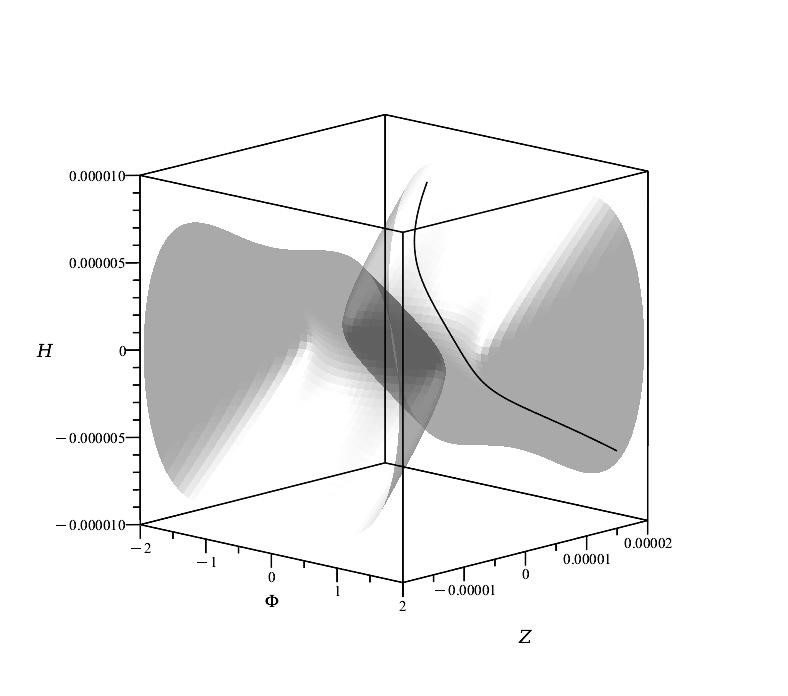}{6}{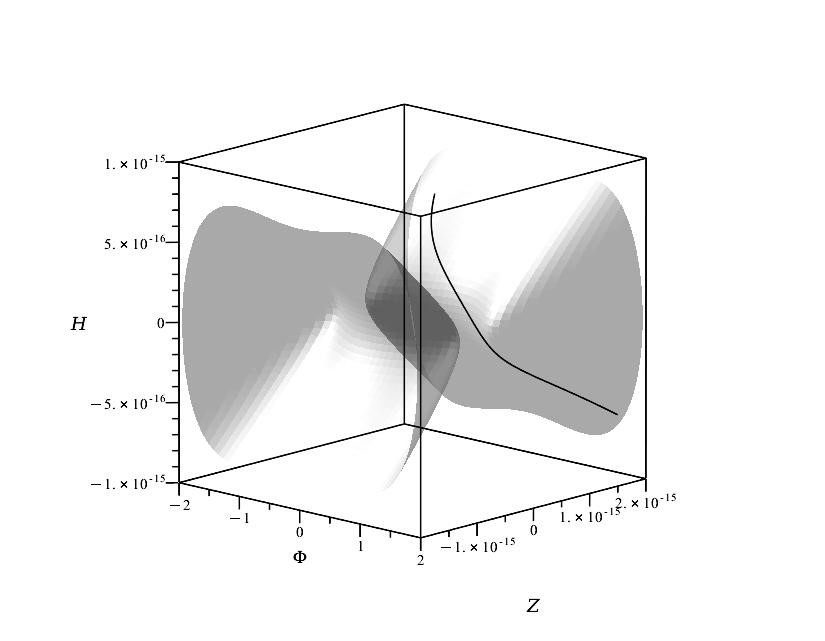}{6}{Phase trajectory of the $\mathbf{M_0}$ model on the surface \eqref{Eq14}\label{Fig10}.}
{Phase trajectory of the model $\mathbf{M_{\mathrm{SU(5)}}}$ on the surface \eqref{Eq14}\label{Fig11}.}
{Phase trajectory of the model $\mathbf{M_{\mathrm{SM}}}$ on the surface \eqref{Eq14}\label{Fig12}.}

Figures \ref{Fig13} - \ref{Fig14} show graphs of the basic functions of the model H(t) and Z(t).
In these graphs, the corresponding functions are presented on time scales corresponding to the transformations \eqref{Eq22},
i.e., for the SU(5) model $\tilde{t}=10^{-5}t$, and for the SM model $\tilde{t}=10^{-15}t$ .
\TwoFigsReg{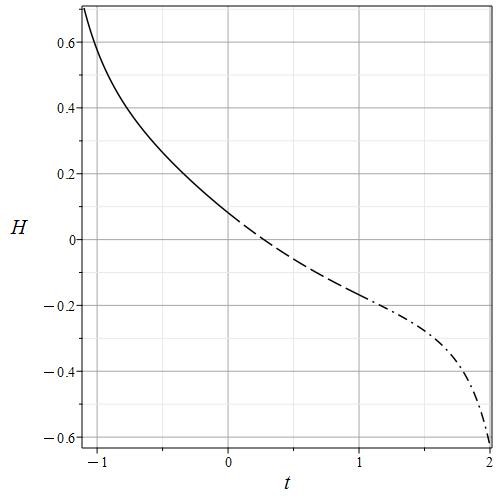}{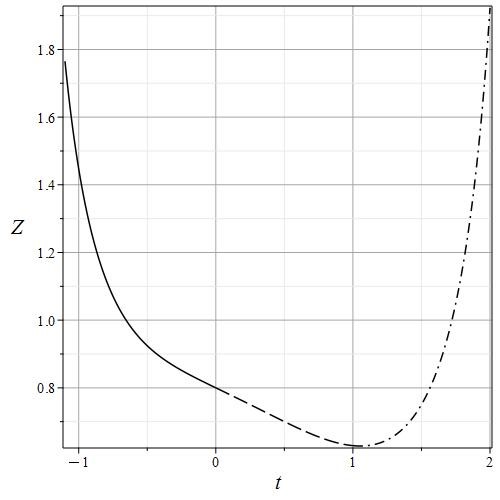}{8}{6}{6}{H(t):solid line - $H_0$;
dashed - $10^5\cdot H_{\mathrm{SU(5)}}$;
dash-dotted - $10^{15}\cdot H_{\mathrm{SM}}$.\label{Fig13}}
{Z(t): solid line - $Z_0$; dashed - $10^{5}\cdot Z_{\mathrm{SU(5)}}$;
  dash-dotted - $10^{15}\cdot Z_{\mathrm{SM}}$.\label{Fig14}}

As can be seen from the presented graphs, they completely obey the law of scaling transformation \eqref{Eq22}, their deviation from the prototype does not exceed $10^{-6}$, i.e., does not exceed the error of numerical integration.
In Fig. \ref{Fig15} shows a plot of the invariant cosmological acceleration $\Omega(t)$ \eqref{Eq18} for these models, confirming invariance (formula (\eqref{Eq23})).
\Fig{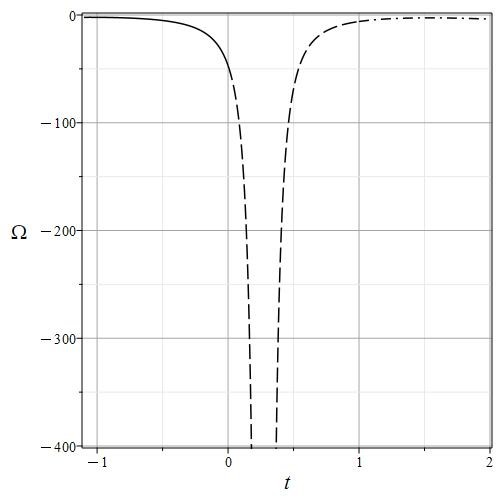}{8}{Invariant cosmological acceleration $\Omega(t)$:
$\mathrm{\mathbf{M_0}}$ - solid line,
$\mathrm{\mathbf{M_{SU(5)}}}$ - dash-dotted line,
$\mathrm{\mathbf{M_{SM}}}$ - dotted point\label{Fig15}}
?
\subsection{$\mathbf{P_2}=[[-1,1],0.1];\mathbf{I_2}=[0.999,0,1]$}
In this case, the Einstein-Higgs surface becomes doubly connected and has two singular points. Figures \ref{Fig16} - \ref{Fig21} show the results of numerical simulations for the $\mathbf{P_2}$ model.
These graphs exhibit similar properties with respect to scaling transformations,
as in the previous case, although the type of trajectory becomes completely different.
As can be seen from the graphs in figures \ref{Fig16} - \ref{Fig18},
phase trajectories of models, as well as their Eistein-Higgs surfaces
obeys the laws of scaling transformation \eqref{Eq22}: in Fig. \ref{Fig17}
the scales of the Z and H axes are stretched $k=10^5$ times, and in Fig. \ref{Fig18} - $k=10^{15}$ times.
\ThreeFigReg{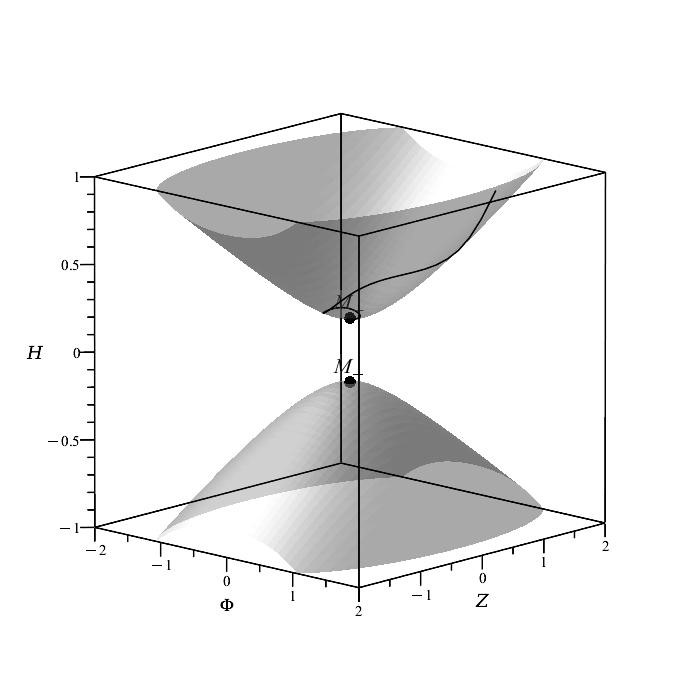}{6}{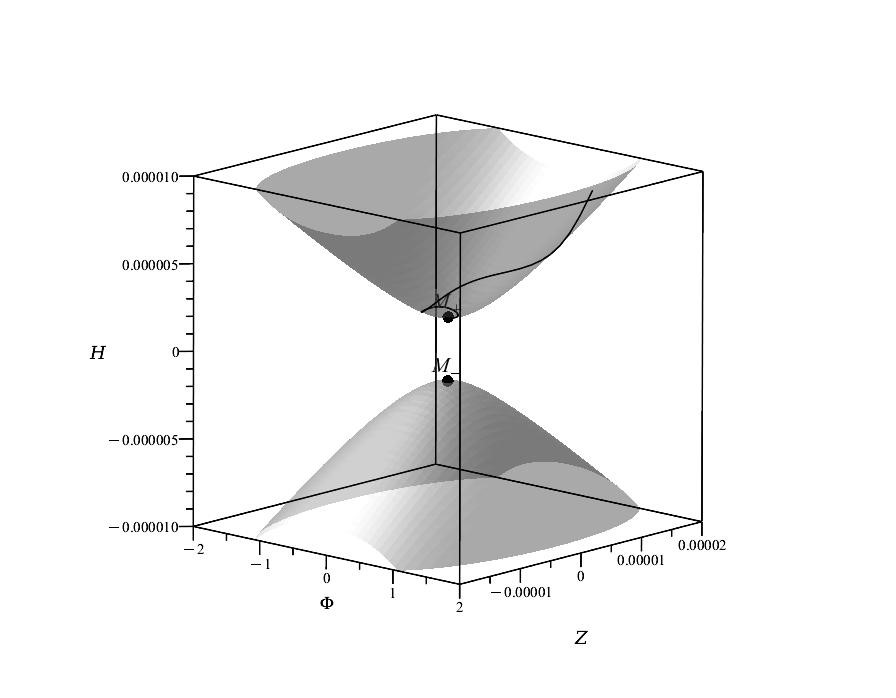}{6}{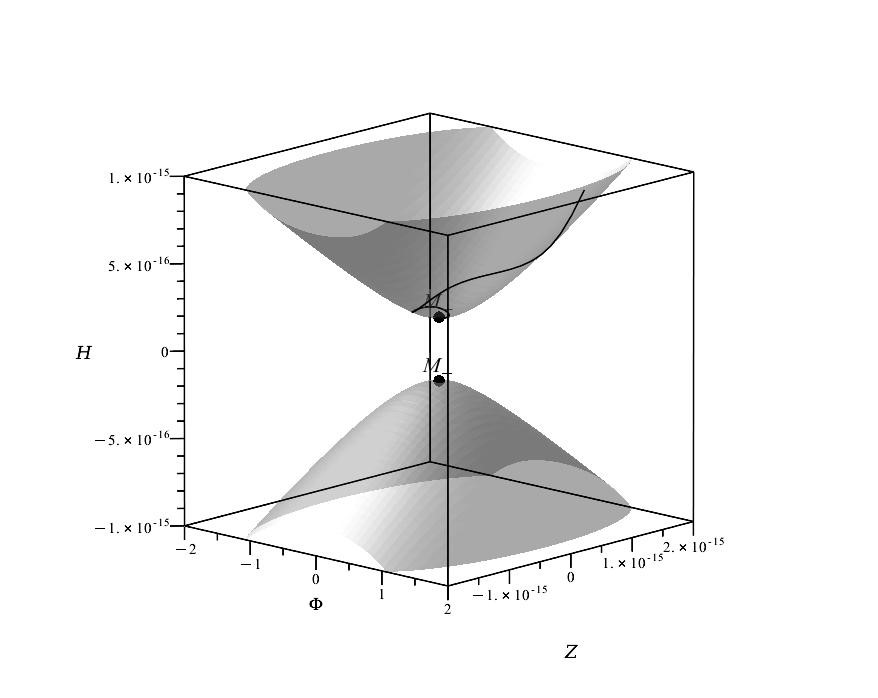}{6}{Phase trajectory of the $M_0$ model on the surface \eqref{Eq14}\label{Fig16}}
{Phase trajectory of the model $M_{\mathrm{SU(5)}}$ on the surface \eqref{Eq14}\label{Fig17}}
{Phase trajectory of the $M_{\mathrm{SM}}$ model on the surface \eqref{Eq14}\label{Fig18}}
Figures \ref{Fig19} - \ref{Fig20} show graphs of the basic functions of the model H(t) and Z(t). In these graphs, the corresponding functions are presented on time scales corresponding to the transformations \eqref{Eq22},
i.e., for the SU(5) model $\tilde{t}=10^{-5}t$ , and for the SM model $\tilde{t}=10^{-15}t$.
\TwoFigsReg{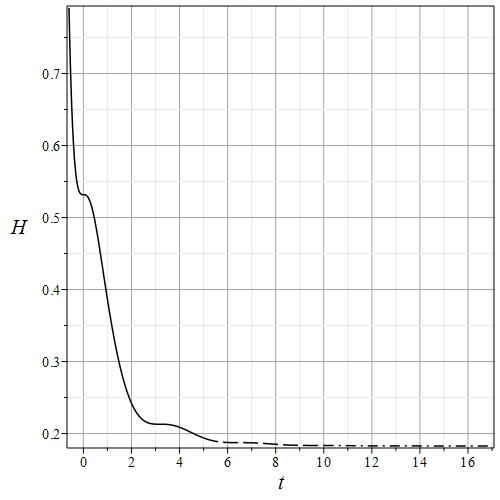}{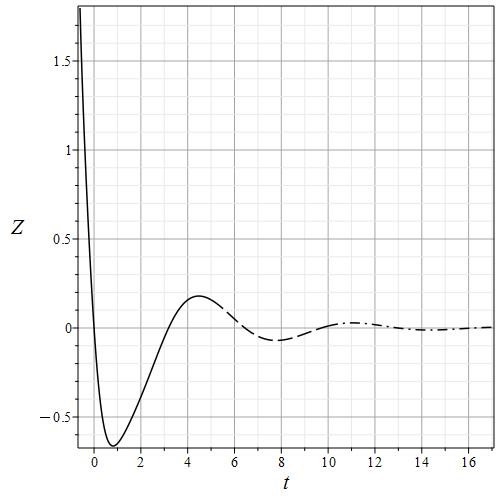}{8}{6}{6}{H(t):solid line - $H_0$;
dashed - $10^5\cdot H_{\mathrm{SU(5)}}$;
dash-dotted - $10^{15}\cdot H_{\mathrm{SM}}$.\label{Fig19}}
{Z(t): solid line - $Z_0$; dashed - $10^{5}\cdot Z_{\mathrm{SU(5)}}$;
  dash-dotted - $10^{15}\cdot Z_{\mathrm{SM}}$.\label{Fig20}}
As can be seen from the presented graphs, they completely obey the law of scaling transformation \eqref{Eq22}, their deviation from the prototype does not exceed $10^{-6}$, i.e., does not exceed the error of numerical integration.
Fig.\ref{Fig21} shows a plot of the invariant cosmological acceleration $\Omega(t)$ \eqref{Eq18} for these models,
confirming e\"e invariance (formula \eqref{Eq23}).
\Fig{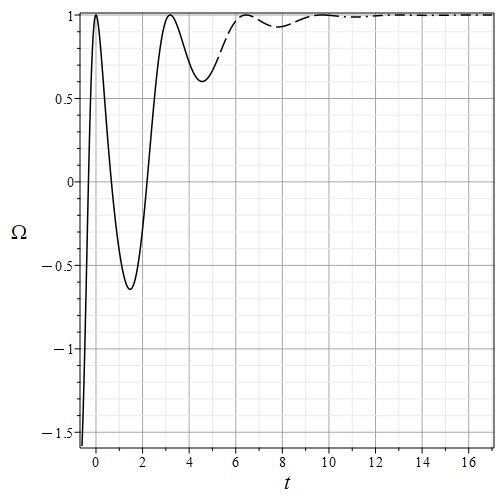}{8}{Invariant cosmological acceleration $\Omega(t)$:
$\mathrm{\mathbf{M_0}}$ - solid line,
$\mathrm{\mathbf{M_{SU(5)}}}$ - dash-dotted line,
$\mathrm{\mathbf{M_{SM}}}$ - dotted point\label{Fig21}}.

\subsection{$\mathbf{P_3}=[[1,1],0.1];\mathbf{I_3}=[0.999,0,1]$}
In this case, the Einstein-Higgs surface becomes simply connected and has six singular points.
Figures \ref{Fig22} - \ref{Fig27} show the results of numerical simulations for the $\mathbf{P_3}$ model.
These graphs exhibit similar properties with respect to scaling transformations,
as in the previous case, although the type of trajectory becomes completely different.
As can be seen from the graphs in figures \ref{Fig22} - \ref{Fig24}, the phase trajectories of the models,
like their Eistein-Higgs surfaces, they obey the laws of scaling transformation \eqref{Eq22}: in Fig. \ref{Fig23}
the scales of the Z and H axes are stretched $k=10^5$ times, and in Fig. \ref{Fig24} - $k=10^{15}$ times.
\ThreeFigReg{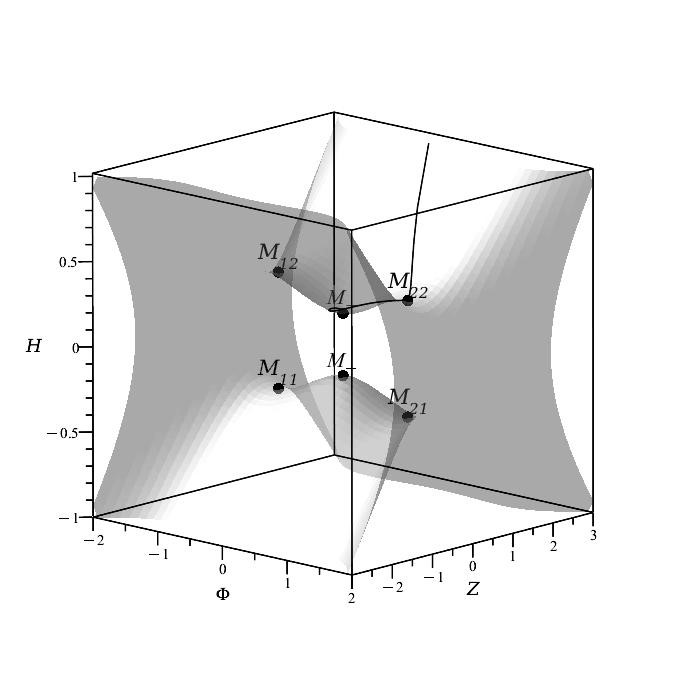}{6}{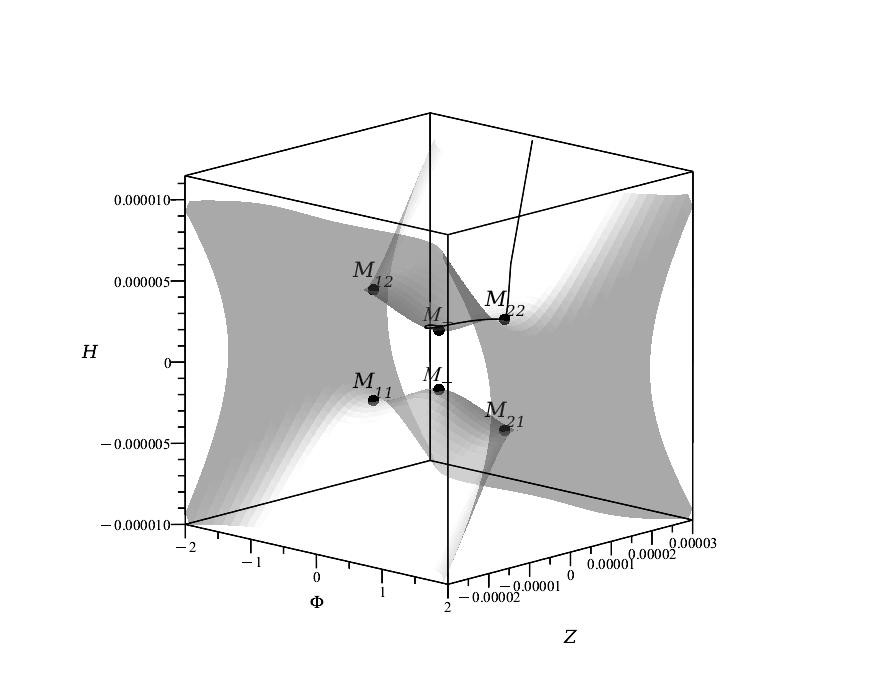}{6}{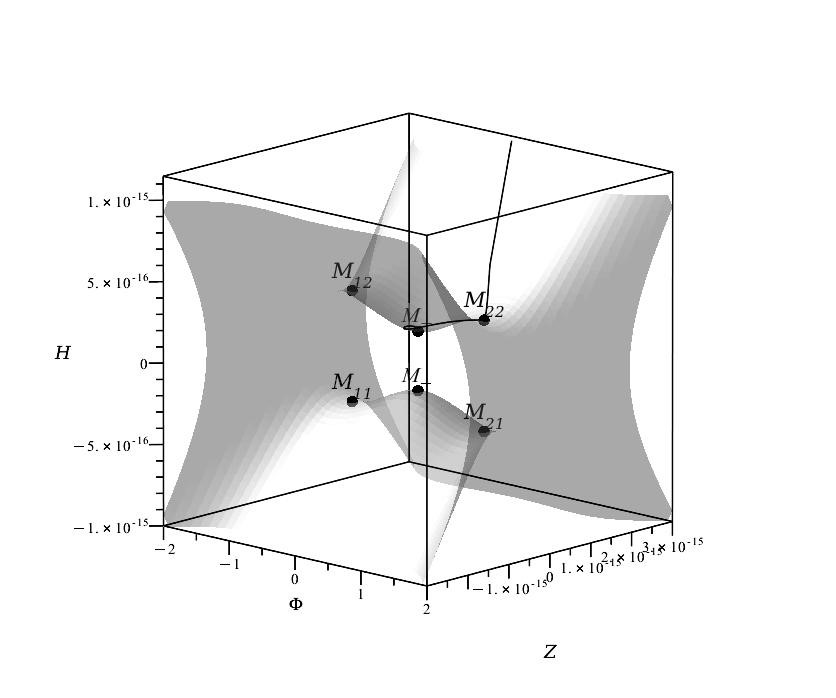}{6}
{Phase trajectory of the model $\mathbf{M_0}$ on the surface \eqref{Eq14}\label{Fig22}.}
{Phase trajectory of the model $\mathbf{M_{\mathrm{SU(5)}}}$ on the surface \eqref{Eq14}\label{Fig23}.}
{Phase trajectory of the model $\mathbf{M_{\mathrm{SM}}}$ on the surface \eqref{Eq14}\label{Fig24}.}
Figures \ref{Fig25} - \ref{Fig26} show graphs of the basic functions of the model H(t) and Z(t). On these graphs, the corresponding functions are presented on time scales corresponding to the transformations \eqref{Eq22}, i.e., for the SU(5) model $\tilde{t}=10^{-5}t$, and for the SM model $\ tilde{t}=10^{-15}t$.

\TwoFigsReg{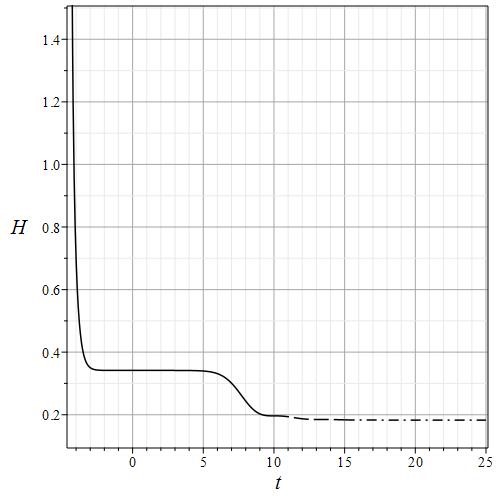}{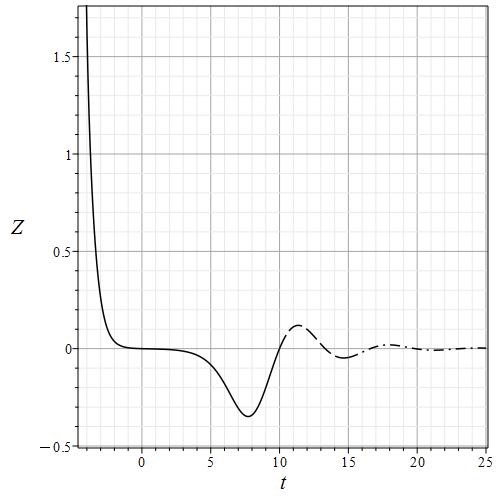}{8}{6}{6}{H(t):solid line - $H_0$;
dashed - $10^5\cdot H_{\mathrm{SU(5)}}$;
dash-dotted - $10^{15}\cdot H_{\mathrm{SM}}$.\label{Fig25}}
{Z(t): solid line - $Z_0$; dashed - $10^{5}\cdot Z_{\mathrm{SU(5)}}$;
  dash-dotted - $10^{15}\cdot Z_{\mathrm{SM}}$.\label{Fig26}}

As can be seen from the presented graphs, they completely obey the law of scaling transformation \eqref{Eq22}, their deviation from the prototype does not exceed $10^{-6}$, i.e., does not exceed the error of numerical integration.
In Fig. \ref{Fig27} shows a graph of the invariant cosmological acceleration $\Omega(t)$ \eqref{Eq18} for these models, confirming e\"e invariance (formula \eqref{Eq23}).
\Fig{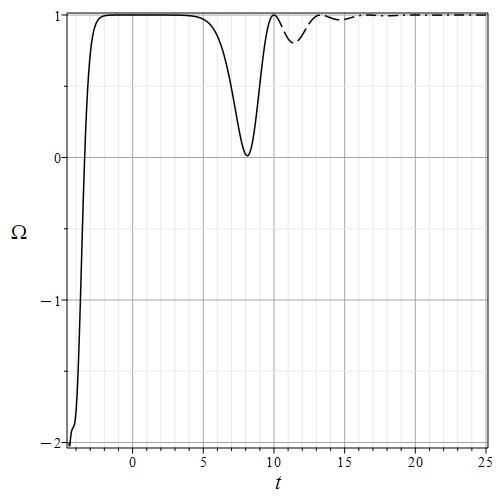}{8}{Invariant cosmological acceleration $\Omega(t)$:
$\mathrm{\mathbf{M_0}}$ - solid line,
$\mathrm{\mathbf{M_{SU(5)}}}$ - dash-dotted line,
$\mathrm{\mathbf{M_{SM}}}$ - dotted point\label{Fig27}.}

\subsection{$\mathbf{P_4}=[[1,1],-0.1];\mathbf{I_4}=[0.999,0,1]$}
In this case, the Einstein-Higgs surface becomes simply connected and has four singular points. Figures \ref{Fig28} - \ref{Fig33} show the results of numerical simulations for the $\mathbf{P_4}$ model. These graphs exhibit similar properties with respect to scale transformations as in the previous case, although the type of trajectory becomes completely different. As can be seen from the graphs in Figures \ref{Fig28} - \ref{Fig30}, the phase trajectories of models, like their Einstein-Higgs surfaces, obey the laws of scaling transformation \eqref{Eq18}:
in Fig. \ref{Fig29} the scales of the Z and H axes are stretched $k=10^{5}$ times, and in Fig. \ref{Fig30} - $k=10^{15}$ times.
\ThreeFigReg{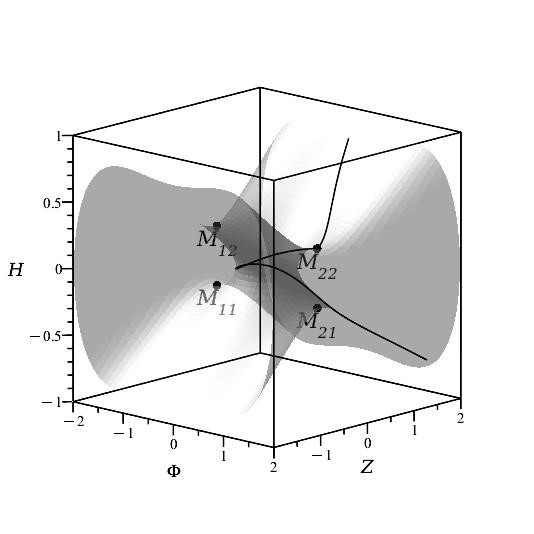}{6}{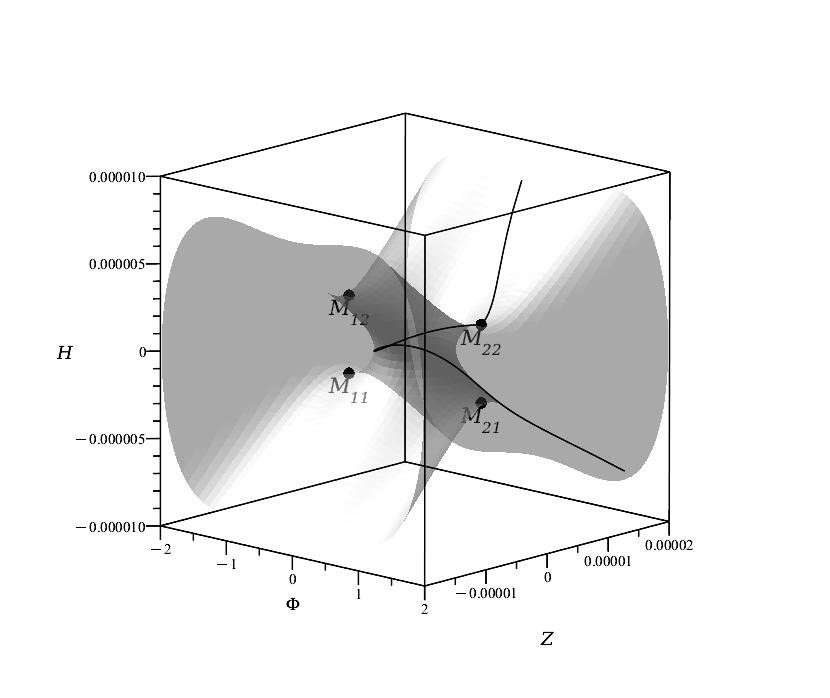}{6}{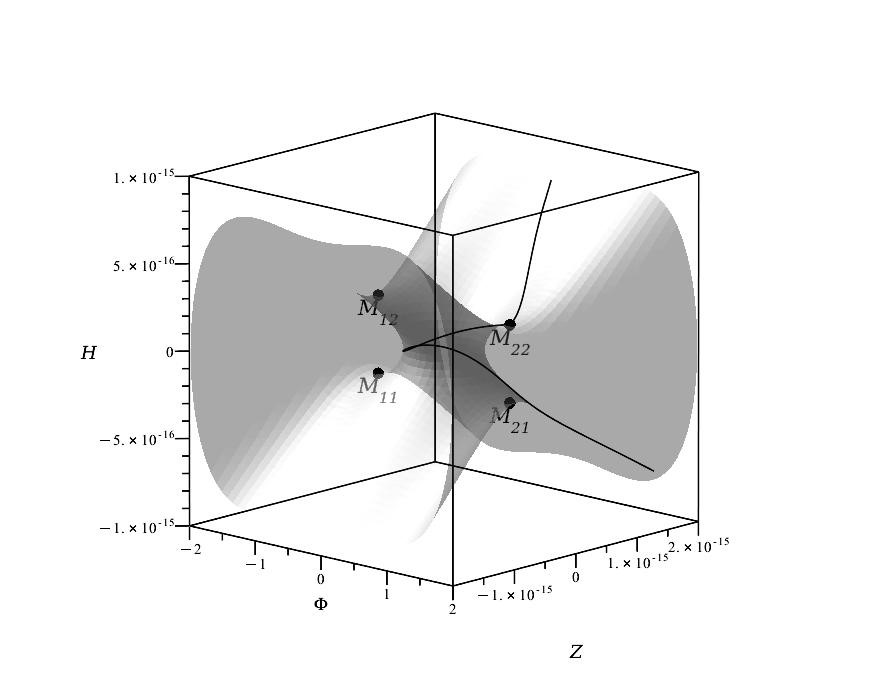}{6}
{Phase trajectory of the model $\mathbf{M_0}$ on the surface \eqref{Eq14}\label{Fig28}.}
{Phase trajectory of the model $\mathbf{M_{\mathrm{SU(5)}}}$ on the surface \eqref{Eq14}\label{Fig29}.}
{Phase trajectory of the model $\mathbf{M_{\mathrm{SM}}}$ on the surface \eqref{Eq14}\label{Fig30}.}

Figures \ref{Fig31} - \ref{Fig32} show graphs of the basic functions of the model H(t) and Z(t). On these graphs, the corresponding functions are presented on time scales corresponding to the transformations \eqref{Eq22}, i.e., for the SU(5) model $\tilde{t}=10^{-5}t$, and for the SM model $\ tilde{t}=10^{-15}t$.

\TwoFigsReg{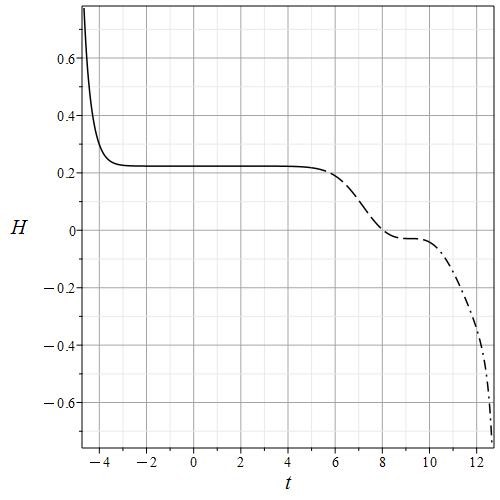}{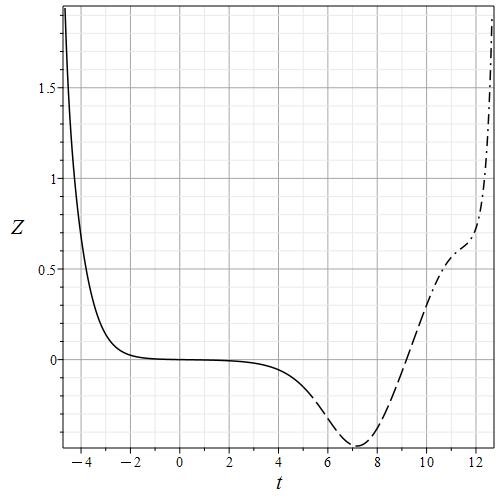}{8}{6}{6}{H(t):solid line - $H_0$;
dashed - $10^5\cdot H_{\mathrm{SU(5)}}$;
dash-dotted - $10^{15}\cdot H_{\mathrm{SM}}$.\label{Fig31}}
{Z(t): solid line - $Z_0$; dashed - $10^{5}\cdot Z_{\mathrm{SU(5)}}$;
  dash-dotted - $10^{15}\cdot Z_{\mathrm{SM}}$.\label{Fig32}}

As can be seen from the presented graphs, they completely obey the law of scaling transformation \eqref{Eq22}, their deviation from the prototype does not exceed $10^{-6}$, i.e., does not exceed the error of numerical integration.
Fig.\ref{Fig33} shows a graph of the invariant cosmological acceleration \eqref{Eq18} for these models, confirming its invariance (formula \eqref{Eq23}). However, note that near the point the errors become significant, which is caused by zero values (see Fig.\ref{Fig31}) in the denominator of the formula\eqref{Eq18}.
\Fig{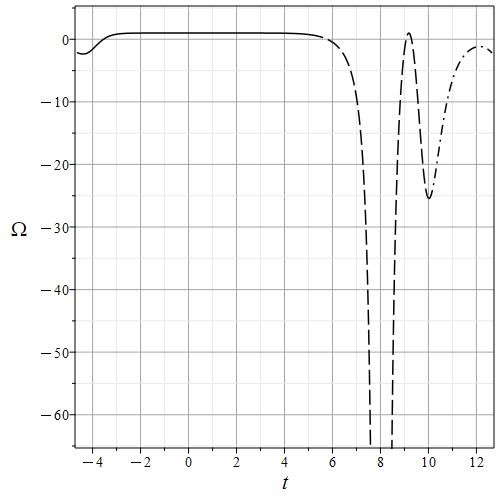}{8}{Invariant cosmological acceleration $\Omega(t)$:
$\mathrm{\mathbf{M_0}}$ - solid line,
$\mathrm{\mathbf{M_{SU(5)}}}$ - dash-dotted line,
$\mathrm{\mathbf{M_{SM}}}$ - dotted point\label{Fig33}.}

\subsection{$\mathbf{P_5}=[[2.5,1],-0.1];\mathbf{I_5}=[0.63,0.0099,1]$}
In this case, the Einstein-Higgs surface becomes a cone and has four singular points. In the figures \ref{Fig34} - \ref{Fig39}
The results of numerical simulation for the $\mathbf{P_5}.$ model are presented.
These graphs exhibit similar properties with respect to scaling transformations,
as in the previous case, although the type of trajectory becomes completely different. As can be seen from the graphs
in the figures \ref{Fig34} - \ref{Fig36}, the phase trajectories of the models, as well as their Einstein-Higgs surfaces
obeys the laws of scaling transformation \eqref{Eq22}: in Fig. \ref{Fig35} the scales of the Z and H axes are stretched $k=10^{5}$ times,
and in Fig. \ref{Fig36} - $k=10^{15}$ times.
\ThreeFigReg{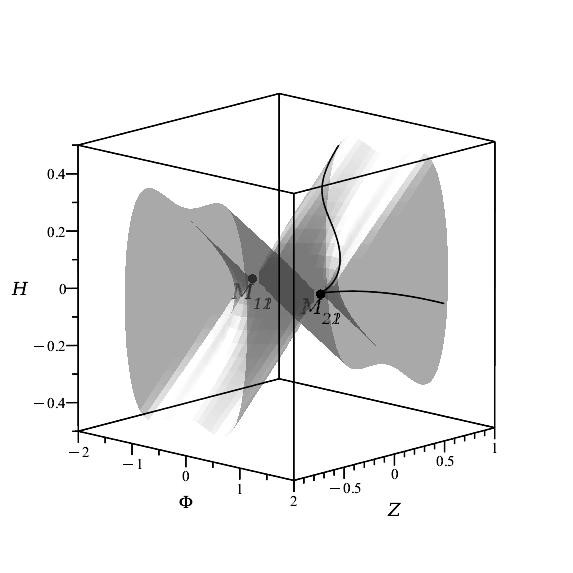}{6}{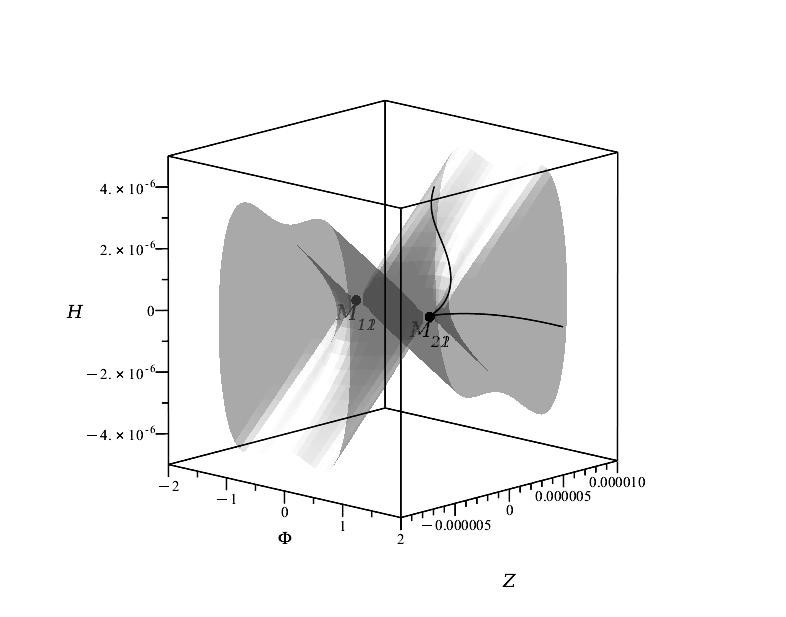}{6}{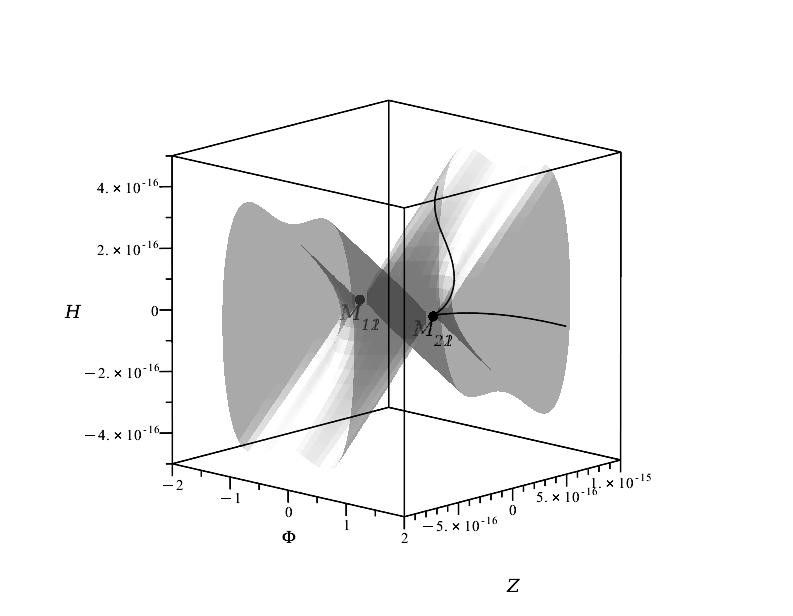}{6}
{Phase trajectory of the model $\mathbf{M_0}$ on the surface \eqref{Eq14}\label{Fig34}.}
{Phase trajectory of the model $\mathbf{M_{\mathrm{SU(5)}}}$ on the surface \eqref{Eq14}\label{Fig35}.}
{Phase trajectory of the model $\mathbf{M_{\mathrm{SM}}}$ on the surface \eqref{Eq14}\label{Fig36}.}

Figures \ref{Fig37} - \ref{Fig38} show graphs of the basic functions of the model H(t) and Z(t).
In these graphs, the corresponding functions are presented on time scales corresponding to
transformations \eqref{Eq22}, i.e., for the SU(5) model $\tilde{t}=10^{-5}t$, and for the SM model $\tilde{t}=10^{-15} t$.
  \TwoFigsReg{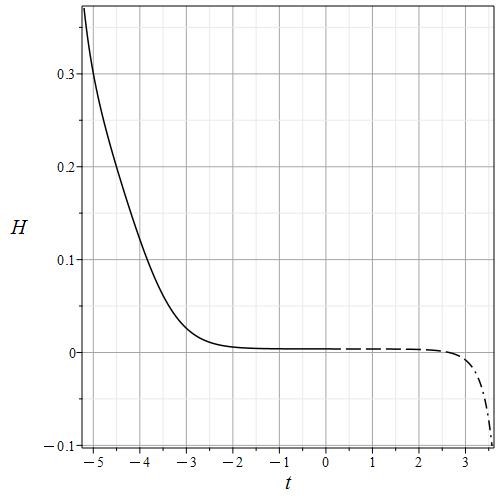}{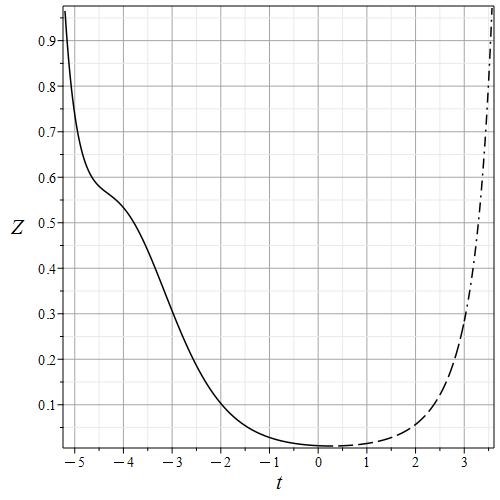}{8}{6}{6}{Fig38}{H(t):solid line - $H_0$;
dashed - $10^5\cdot H_{\mathrm{SU(5)}}$;
dash-dotted - $10^{15}\cdot H_{\mathrm{SM}}$.\label{Fig37}}
{Z(t): solid line - $Z_0$; dashed - $10^{5}\cdot Z_{\mathrm{SU(5)}}$;
  dash-dotted - $10^{15}\cdot Z_{\mathrm{SM}}$.\label{Fig38}}
As can be seen from the presented graphs, they completely obey the law of scaling transformation \eqref{Eq22}, their deviation from the prototype does not exceed $10^{-6}$, i.e., does not exceed the error of numerical integration.
Fig.\ref{Fig39} shows a graph of the invariant cosmological acceleration $\Omega(t)$ \eqref{Eq18} for these models, confirming e\"e invariance (formula \eqref{Eq23}). However, note that near the error points become significant, which is caused by the zero values (see Fig. \ref{Fig37}) in the denominator of the formula \eqref{Eq18}.
\Fig{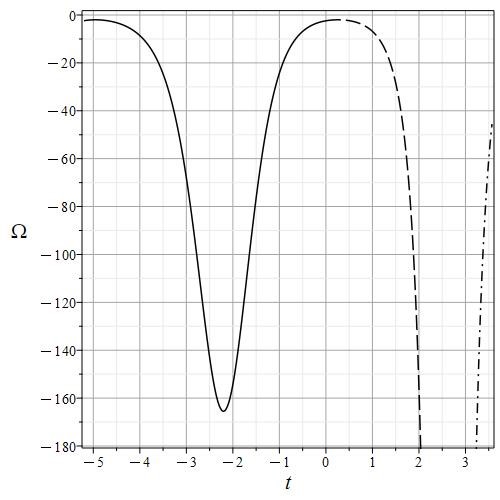}{8}{Invariant cosmological acceleration $\Omega(t)$:
$\mathrm{\mathbf{M_0}}$ - solid line,
$\mathrm{\mathbf{M_{SU(5)}}}$ - dash-dotted line,
$\mathrm{\mathbf{M_{SM}}}$ - dotted point\label{Fig39}.}

\section{Conclusion}

To summarize the research, let us briefly summarize its main results.
\begin{enumerate}
\item The similarity properties of cosmological models based on a classical scalar field with the Higgs potential are formulated and proven.
\item The indicated properties are confirmed by numerical modeling with high integration accuracy.
\item The similarity properties of cosmological models on different physical scales of field-theoretic interaction models have been revealed: Planck level models, SU(5) level models and SM level - Standard Model.
\item The formulated properties of similarity make it possible to quite simply predict the behavior of cosmological models built on various field-theoretic models of interaction.
\end{enumerate}
	
\textbf{Funding}

The work is carried out in accordance with the Strategic Academic Leadership Program "Priority 2030" of the Kazan Federal University of the Government of the Russian Federation.


\end{document}